\documentclass[print]{aa}
\usepackage{amsmath,amssymb}
\usepackage{graphicx}
\usepackage{hyperref}

\newcommand{\fig}[1]{Fig. \ref{#1}}
\newcommand{\refsection}[1]{Section \ref{#1}}

\newcommand{\refeq}[1]{Eq. (\ref{#1})}
\newcommand{\refeqs}[2]{Eqs. (\ref{#1} - \ref{#2})}

\newcommand{\BothTen}{DM10G10}
\newcommand{\BothPointOne}{DM0.1G0.1}
\newcommand{\BothOne}{DM1G1}
\newcommand{\DMTen}{DM10G1}
\newcommand{\DMPointOne}{DM0.1G1}
\newcommand{\GasTen}{DM1G10}
\newcommand{\GasPointOne}{DM1G0.1}

\begin{document}

\title{Cosmological simulations with hydrodynamics of screened scalar-tensor gravity with non-universal coupling}
\titlerunning{Non-universal coupling in screened scalar-tensor gravity}
\author{A. Hammami
        \inst{1}
        \and
        D. F.  Mota
        \inst{1}}

\institute{Institute of Theoretical Astrophysics, University of Oslo, P.O. Box 1029 Blindern, N-0315 Oslo, Norway\\
            \email{amirham@astro.uio.no}\\ 
            \email{d.f.mota@astro.uio.no}}

\abstract
{}
{We study the effects of letting dark matter and gas in the Universe couple to the scalar field of the symmetron model, a modified gravity theory, with varying coupling strength. We also search for a way to distinguish between universal and non-universal couplings in observations.}
{The research is performed utilising a series of hydrodynamic, cosmological N-Body simulations, studying the resulting power spectra and galaxy halo properties, such as  density and temperature profiles.}
{In the cases of universal couplings, the deviations in the baryon fraction from $\Lambda$CDM are smaller than in the cases of non-universal couplings throughout the halos. The same is apparent in the power spectrum baryon bias, defined as the ratio of gas to dark matter power spectrum. Deviations of the density profiles and power spectra from the $\Lambda$CDM reference values can differ significantly between dark matter and gas because the dark matter deviations  are mostly larger than the deviations in the gas.}
{}

\maketitle

\section{Introduction}
One of the most challenging problems in the field of cosmology is to understand the accelerated expansion of the late-time Universe \citep{acc_universe}. The $\Lambda$CDM model is the most accepted explanation for this expansion and is reached by modifying general relativity (GR) through adding a cosmological constant, dark energy, to the energy-momentum tensor. 

An alternative to adding a dark energy component to GR is to modify gravity, by altering the Einstein-Hilbert Lagrangian, which  the Einstein tensor is derived from. Many of these modified gravity theories exist \citep{BransDick,f(R)Theories,Sotiriou:2006hs,mod_grav_review,Boehmer} and several have previously been studied \citep{nbodycham,Barreira,Li1,Li2,Davis,Barrow,Silk,MG-gadget,Winther}. These theories are often implemented by introducing a scalar field to the Einstein tensor, which  couples to the matter component of the Universe. This scalar field gives rise to a fifth force, an additional gravitational force, which is negligible
at solar system scales and below, according to laboratory experiments \citep{labexp,labexp2} and solar system gravity probes \citep{solarsyst, cwillreview}. 

If we assume that this fifth force acts on larger scales than the solar system, then some mechanism is needed to negate the fifth force on solar system scales. One way to achieve this is to utilise one of the screening mechanisms found in the literature \citep{screen1, screen2, SymmetronPaper,Chameleons,Vainshtein,Koivisto} that screens the fifth force  based on a series of different criteria. In this paper, we  study the symmetron model \citep{SymmetronPaper},  which screens the fifth force  in regions of high density. 

With these modified gravity theories comes the challenge of finding methods to test them against observations \citep{TestCham, XMM}. Theorists in the past have mainly used predictions from  models
 and simulations that only include dark matter due to the simplistic nature of  dark matter, when constraining modified gravity theories. Use of  models that only include dark matter can be justified by the matter composition of the Universe, which is 84.4\% dark matter and 15.6\% baryonic matter, according to the \citet{Planck}. However, astronomers observe the electromagnetic spectrum emitted from the baryonic components of the Universe, leaving the community with a disconnect between theories and observables.

Introducing the concept of a bias between the dark matter and baryonic components is one way to rectify this disconnect. The bias assumes that the behaviour of the two components are the same, but that the strength or amplitude of the behaviour might be different. The community generally assumes that the bias of the standard model $\Lambda$CDM is equal to one, which justifies  studying the Universe and comparing the observations to  simulations that only include dark matter. However, if the bias is not unity researchers might greatly underestimate or overestimate their findings. 

The bias can deviate from unity even in $\Lambda$CDM because of baryonic effects other than a non-universal coupling, such as AGN-feedback \citep{AGN1} or the hydrodynamic effect observed in the Bullet cluster \citep{bullet_cluster, clowe}. This means that these baryonic processes need to be comprehensively understood before the bias can be used to test a non-universal coupling.
Recent work \citep{newpaper} has made progress in studying the differences between galaxies (gas) and dark matter in the $\Lambda$CDM model.

In our previous work \citep{hydro_mod}, we studied the effects of adding a hydrodynamic gas to an existing N-body code with the symmetron model implemented \citep{ISIS} with the  assumption that the scalar field would couple to the gas and the dark matter to a universal coupling. There are, however, no justifications for assuming that the dark matter and gas have the same coupling to the scalar field. In fact, several known particles, such as neutrinos and baryons, do not interact in the same manner with electromagnetic forces, and it therefore stands to reason that it is worthwhile to study the effects of  non-universal coupling between the scalar field and  matter components.

In this paper, we  study  density and temperature profiles and power spectra of dark matter and gas for  symmetron models with different values for the coupling strength. The paper starts with an introduction to the symmetron model in \refsection{scaltenstheory}, followed by a very brief section on the simulation parameters in \refsection{code_implem}, presenting and discussing the power spectra in \refsection{power_results} and the density and temperature profiles in \refsection{profile_results}, and finishes with conclusions in \refsection{discuss}.

We work with some models that are extremely coupled to the scalar field to push any signatures of the non-universal coupling to their limits and possibly reveal signatures that would not  immediately be clear from more sensible scalar field couplings.

\section{The symmetron model}
\label{scaltenstheory}
Introduced by \citet{SymmetronPaper} the symmetron model is a scalar theory of gravity using a symmetric potential, where the action \citep{Sotiriou:2006hs,2003CQGra..20.4503F} is
\begin{align}
 S &= \int d^4x\sqrt{-g}\left[\frac{R}{2}M_{\rm pl}^2 - \frac{1}{2}\partial^i\psi\partial_i\psi - V(\psi)\right] \label{Sym_action} \\
 &\quad+ S_m(\tilde{g}_{\mu\nu},\tilde{\Psi}_i),  \nonumber
\end{align}
where $\psi$ is the scalar field\footnote{We only study the quasi-static limit \citep{2013PhRvL.110p1101L, quasistatic} of the scalar field, where time derivatives are ignored.}, $R$ is the Ricci scalar, $M_{\rm pl}$ is the Planck mass, and $g=|g_{\mu\nu}|$ is the determinant of the metric tensor in the Einstein frame, which can be converted to the Jordan frame by
\begin{align}\label{conftrans}
 \tilde{g}_{\mu\nu} = A^2(\psi)g_{\mu\nu}.
\end{align}
The conformal factor satisfies $A \simeq 1$ for the symmetron model and we use this approximation throughout. For more on these frames and the transformations between them and possible errors, see \citet{Faraoni:1998qx} and \citet{BrownHammami}.

To preserve the behaviour of gravity, as described by GR,  the symmetron model utilises a screening mechanism that triggers based on a set density value $\rho_{\rm assb}$ at the solar system
scale(a region of high density). In regions of low density, the symmetron would produce a modification of order one on the gravity. To accomplish this, the potential in the action above is defined to be symmetric, as in 
\begin{align}
V(\psi) = V_0-\frac{1}{2}\mu^2\psi^2 + \frac{1}{4}\lambda\psi^4,
\end{align}
where $\psi$ is the scalar field, $\mu$ is a mass scale, and $\lambda$ a dimensionless parameter. Likewise, the coupling factor is also symmetric, 
\[A(\psi) = 1 + \frac{1}{2}\left(\frac{\psi}{M}\right)^2, \]
with $M$ being another mass scale.

To find the stress energy tensor for the symmetron model, we vary the action with respect to the metric
\begin{align}
 T_{\mu\nu} &= A(\psi)T_{\mu\nu}^{(m)} + T_{\mu\nu}^{(\psi)}\nonumber\\
&= A(\psi)\left[(P+\rho)u_{\mu}u_{\nu} + Pg_{\mu\nu}\right]  \label{Sym_stress} \\
&\quad+ \nabla_{\mu}\psi\nabla_{\nu}\psi - g_{\mu\nu}\left(\frac{1}{2}\partial^i\psi\partial_i\psi + V(\psi)\right), \nonumber 
\end{align}
where $P$ and $\rho$ are the pressure and density, respectively. 

The scalar field component of the stress energy tensor is not covariantly conserved, i.e.
\[\nabla^{\nu}T_{\mu\nu}^{(\psi)}\neq0,\] 
while the total stress energy tensor is \citep{Gravitation}
\begin{align}
\nabla^{\nu}T_{\mu\nu}=0.\label{cons_law}
\end{align}

The equation of motion for the scalar field is found by varying the action again, this time with respect to the scalar field,
\begin{align}
 \Box\psi = V'(\psi) - A'(\psi) T^{(m)}, \label{Sym_EoM}
\end{align}
where $T^{(m)}$ is the trace of the stress energy tensor ${T^{(m)}=g^{\mu\nu}T^{(m)}_{\mu\nu}}$. 

The right  side of \refeq{Sym_EoM} can be recognised as an effective potential, and, using \refeq{Sym_stress}, we write\begin{align}
V_{\rm eff}(\psi) = V_0 + \frac{1}{2}\left(\frac{\rho_m}{M^2}-\mu^2\right)\psi^2 + \frac{1}{4}\lambda\psi^4.
\end{align}

With this potential, the scalar field goes to zero in regions of high density, $\rho_m \gg M^2\mu^2$, while in regions of low density it reaches a minimum of $\psi_0 = \pm\mu\sqrt{\frac{1}{\lambda}}$. The addition to gravity, the fifth force scales with the value of the scalar field, and we see from this that it is suppressed in regions of high density.

We redefine the free parameters  $\mu$, $M,$ and $\lambda$ to $\beta$, $\lambda_0,$ and $a_{\rm SSB}$ to a set of parameters that are more physical intuitive as described in \citet{Winther},
\begin{align}
\label{sym_params}
\beta  &= \frac{M_{\rm pl}\psi_0}{M^2}, \\
a_{\rm SSB}^3 &= \frac{3H_0^2\Omega_m M_{\rm pl}^2}{M^2\mu^2}, \\
\lambda_0^2 &= \frac{1}{2\mu^2},
\end{align}
where $H_0$ is the Hubble factor at present day (z=0) and $\Omega_m$ is the matter density parameter.

The relative strength of the fifth force to the gravitational force is represented by $\beta$, the moment of  breaking symmetry is represented by the expansion factor $a_{\rm SSB}=\left(\Omega_{\rm m0}\rho_{\rm c0}/\rho_{\rm SSB}\right)^{1/3}$, and the range of the fifth force is represented by $\lambda_0$ in units of Mpc/h. 

A dimensionless scalar field $\chi$ is defined as
\begin{align}
 \chi &\equiv \frac{\psi}{\psi_0},
\end{align}
with an equation of motion in the quasi-static limit \citep{ISIS}\footnote{Simulations beyond the static limit were presented in \citet{2013PhRvL.110p1101L, nbodysymmnqs}, finding only sub-percent differences between the static and non-static solutions.} as
\begin{align}
 \nabla^2\chi = \frac{a^2}{2\lambda_0}\left[\left(\frac{a_{\rm SSB}}{a}\right)^3\frac{\rho_m}{\overline{\rho}_m}\chi + \chi^3 - \chi\right],
\end{align}
where $\overline{\rho}_m$ is the mean density.

The equation of motion for the position $x$ of the dark matter N-Body particles has been derived from \refeq{Sym_action} in \citet{ISIS} and takes the form
\begin{align}
 \ddot{x} + 2H\dot{x} + \frac{1}{a^2}\nabla\Phi + \frac{1}{a^2}\frac{A'(\psi)}{A(\psi)}\nabla\psi = 0,
\end{align}
where $\Phi$ is the Newtonian gravitational potential.

The fluid equations for the symmetron model is a special case of the general fluid equations for a scalar-tensor theory. Via  the action \refeq{Sym_action}, stress-energy tensor \refeq{Sym_stress}, conservation law \refeq{cons_law}, and  working in the Newtonian Gauge,
\begin{align}
 ds^2 = -(1+2\Phi)dt^2 + a^2(1-2\Phi)\delta_{ij}dx^idx^j,
\end{align}
the fluid equations are derived,
\begin{align}
&\frac{\partial \rho}{\partial t} + \nabla(v\rho)  +3H\rho = 0,\label{eq1}\\
a^2(P+\rho)\Big[Hv &+ \frac{\partial v}{\partial t} + (v\cdot\nabla) v + \frac{1}{a^2}\nabla\Phi\Big] \\
+ \nabla P &+ \frac{A'(\psi)}{A(\psi)}\rho\nabla\psi = 0, \nonumber \\
&\frac{\partial E}{\partial t} + 2HE + v\cdot\nabla E + \frac{P}{\rho}\cdot\nabla v \label{eq3} \\
 = &- (v\cdot\nabla)\Phi - \frac{A'(\psi)}{A(\psi)} (v\cdot\nabla)\psi \nonumber,
\end{align}
where $H=\frac{\dot{a}}{a}$ is the Hubble factor, $v$ is the velocity of the fluid, and $E$ is the internal energy of the fluid.

To remove explicit dependencies on $a$ and $H$ from the equations above, we use a variation of the super-comoving coordinates from \citet{SuperCom}, represented by a tilde,
\begin{align}
\tilde{\chi}=a\chi,\qquad  d\tilde{t} = a^{-2}dt,\qquad  \tilde{\rho} = a^3\rho,\qquad  \tilde{v} = a^2v, \\
 \tilde{\psi} = a\psi, \qquad \tilde{P} = a^5P,\qquad  \tilde{\Phi} = a^2\Phi, \qquad \tilde{E} = a^2E;
\end{align}
 all equations from this point on are in comoving coordinates.

By excluding terms of second order and assuming static pressure, the field \refeqs{eq1}{eq3} transform to\footnote{With this transformation, the derivative in $A'(\tilde{\psi})$ is now with respect to $\tilde{\psi}$.}

\begin{align}
\frac{\partial \tilde{\rho}}{\partial \tilde{t}} + \nabla(\tilde{v}\tilde{\rho}) &= 0,\label{final_cont}\\
\frac{\partial \tilde{v}}{\partial \tilde{t}} + (\tilde{v}\cdot\nabla)\tilde{v} = - \frac{1}{\tilde{\rho}}\nabla \tilde{P} &-\nabla\tilde{\Phi} - \frac{A'(\tilde{\psi})}{A(\tilde{\psi})}\nabla\tilde{\psi} \label{final_euler}, \\
\frac{\partial \tilde{E}}{\partial \tilde{t}} + \tilde{v}\cdot\nabla\tilde{E} + \frac{\tilde{P}}{\tilde{\rho}}\cdot\nabla \tilde{v} = - (\tilde{v}\cdot&\nabla)\tilde{\Phi} - \frac{A'(\tilde{\psi})}{A(\tilde{\psi})} \tilde{v}\cdot\nabla\tilde{\psi}.\label{final_energy}
\end{align}

With this approach, the symmetron model version of the fifth force is
\begin{align}
F_{\psi} &= -\frac{A'(\tilde{\psi})}{A(\tilde{\psi})}\nabla\tilde{\psi} = -\frac{\dfrac{\tilde{\psi}}{M^2}}{1 + \frac{1}{2}\left(\!\dfrac{\tilde{\psi}}{M}\!\right)^2}\nabla\tilde{\psi} \approx -\frac{\tilde{\psi}}{M^2}\nabla\tilde{\psi}\nonumber \\
&= -6\Omega_m H_0^2\frac{(\beta\lambda_0)^2}{a_{\rm SSB}^3}\tilde{\chi}\nabla\tilde{\chi}.
\end{align}
For more on the symmetron model, see \citet{SymmetronPaper}.

\section{Parameters}
\label{code_implem}

The coupling factor defined above is split into two new coupling factors, 
\begin{align}
 \beta \to \begin{cases} \beta_{\rm DM} \\ \beta_{\rm Gas}\end{cases},
\end{align}
which replace the coupling factor in the dark matter and  fluid equations, respectively.

In order to study the effect, we choose couplings of varying orders of magnitude\footnote{The extreme couplings with $\beta=10$ can induce accelerations upwards of $F_{\psi} = 200F_{\rm GR}$. Accelerations of this kind can result in relativistic velocities, requiring a relativistic set of equations to properly describe the systems. Luckily, none of our models induced relativistic velocities, the fastest dark matter particle in our simulations reached a speed of $v_{\rm max}=0.033$c for the \BothTen{} model, barely a factor of 2 higher than the fastest dark matter particle in the $\Lambda$CDM model with $v_{\rm max}=0.016$c. We continue with the extreme models  to push any signatures of non-universal coupling to its limit.} and our chosen configurations are shown in Table 1.

\begin{table}
 \begin{center}
\caption{Coupling factor combinations explored.\label{tab:mgparam}}
  \begin{tabular}{lrr}\hline
  Configuration & $\beta_{\rm DM}$ & $\beta_{\rm Gas}$ \\ \hline
  \BothOne & 1.0 & 1.0\\
  \BothTen & 10.0 & 10.0\\
  \BothPointOne & 0.1 & 0.1\\
  \DMTen & 10.0 & 1.0\\
  \GasTen & 1.0 & 10.0\\
  \DMPointOne & 0.1 & 1.0\\
  \GasPointOne & 1.0 & 0.1\\
  \end{tabular}
 \end{center}
\end{table}

The simulations were run using 1024 cores, $256^3$ dark matter particles, with a box width of $256$ Mpc/h, and six levels of refinements. The background cosmology is a standard 
$\Lambda$CDM background, with $h=0.65$, $\Omega_{\Lambda} = 0.65$, $\Omega_m=0.35,$ and $\Omega_b = 0.05$. The chosen symmetron model has $a_{\rm SSB}=0.33$ and $\lambda_0=1$ Mpc/h.

Because of the use of  extremely coupled models, we ran a set of convergence tests  to verify that the errors induced by  extreme coupling were not too extensive. The tests showed that the code handled the extreme models well for the most part, however,  for the power spectra there were slight errors at the smallest scales $k>4$ Mpc/h and for the halo profiles at radius above  $R>3R_{\rm 200c}$. These errors were not significantly large, but results from the extreme models should be taken with a grain of salt in these regions.

All results in the following sections only focus  on the present day epoch, which corresponds to $z=0$.

\section{Power spectra}
\label{power_results}
With the use of the open POWMES code \citep{POWMES}, we compute the power spectra for both dark matter and gas. To calculate the gas power spectrum, we treat each cell as a particle with a mass defined as
\begin{align}
 m = \rho V_{\rm cell,}
\end{align}
where $\rho$ and $V_{\rm cell}$ is the gas density and volume of the cell, respectively.

\begin{figure*}
        \centering
        \centering
        Dark matter \hspace{5.5cm}\;Gas\\
        \vspace{-2mm}
          \includegraphics[width=0.4\textwidth]{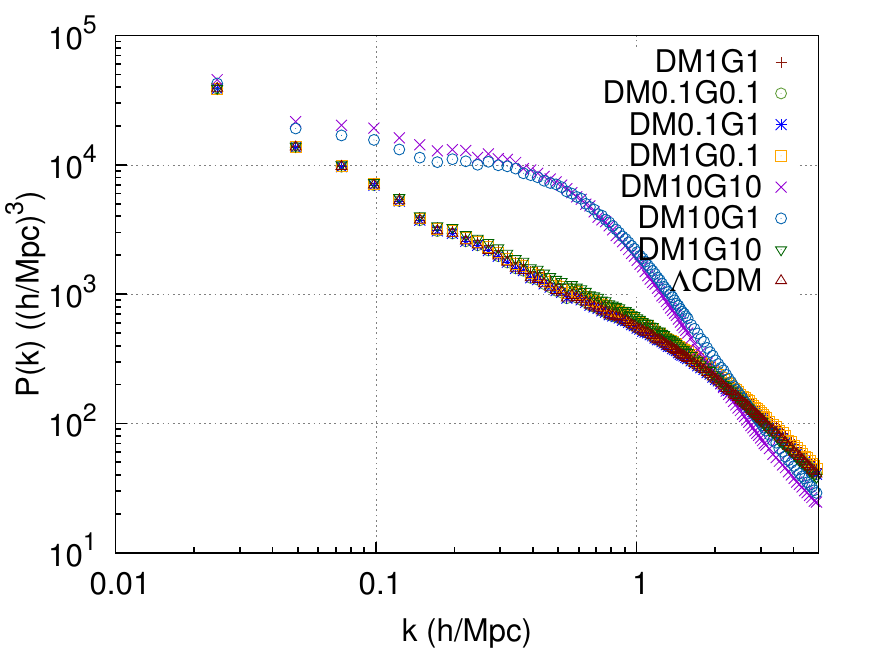}
        \vspace{-3 mm}
        \hspace{-5 mm}
        \includegraphics[width=0.4\textwidth]{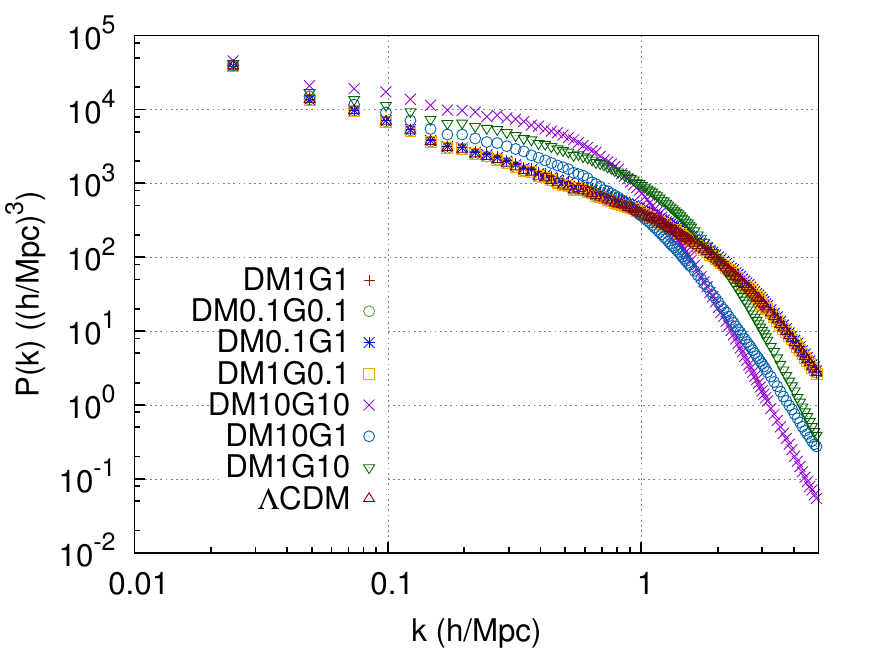}

        \includegraphics[width=0.4\textwidth]{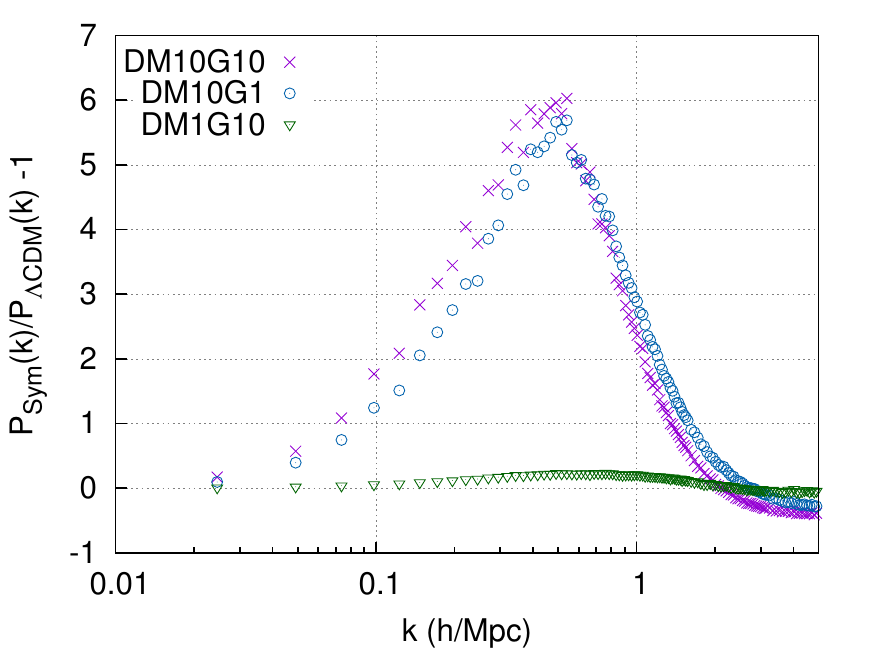}
        \vspace{-3 mm}
        \hspace{-5 mm}
        \includegraphics[width=0.4\textwidth]{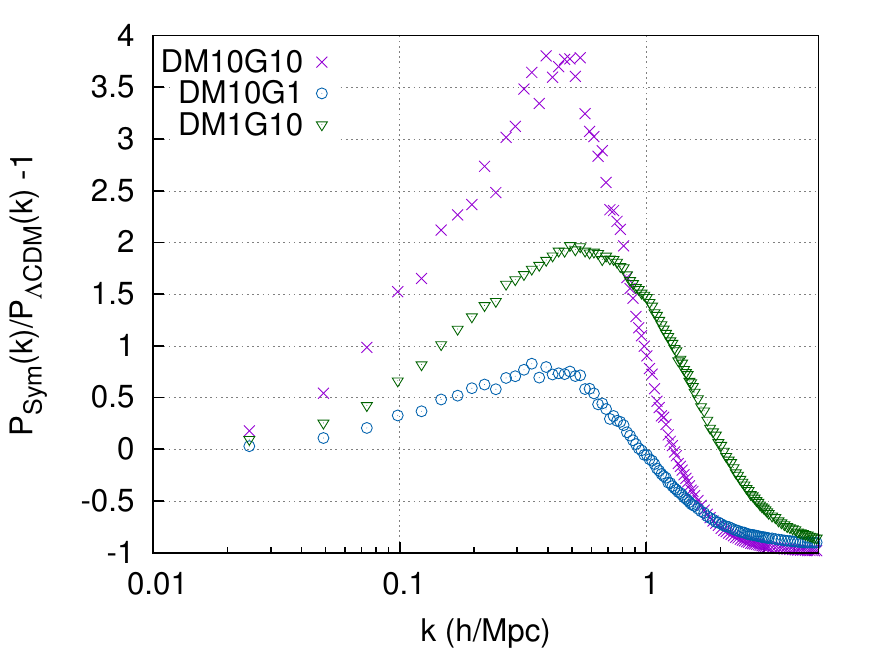}

        \includegraphics[width=0.4\textwidth]{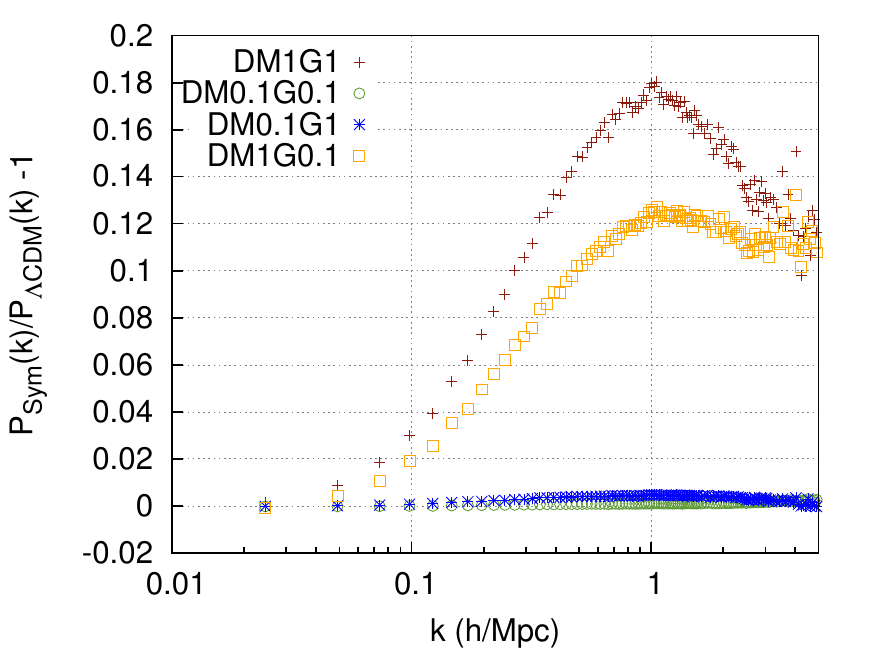}
        \vspace{-3 mm}
        \hspace{-5 mm}
        \includegraphics[width=0.4\textwidth]{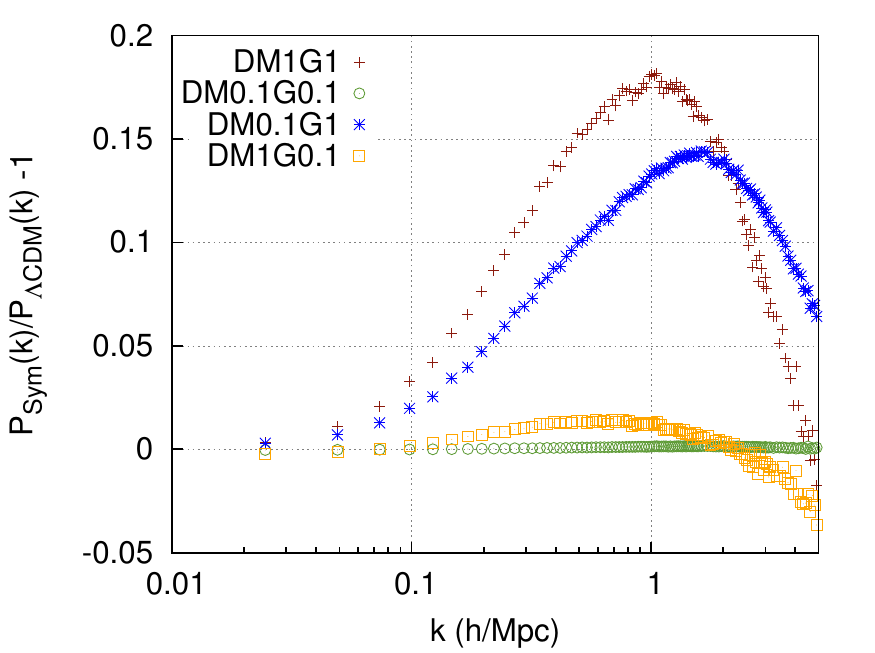}

 \caption{Top: power spectra for all our models. Middle: power spectra deviations of our extreme models from the $\Lambda$CDM model. Bottom: power spectra deviations for the remaining models. Left column shows the dark matter component, while the right column shows the gas component. All results are at $z=0$.}\label{fig:Symmetron_Powerspectrums}
\end{figure*}

In \fig{fig:Symmetron_Powerspectrums} we present the power spectra for all our models and the deviations of these power spectra from $\Lambda$CDM for both the dark matter and gas components.

At the large scale range of the power spectra, $k<1$ Mpc/h, the extreme dark matter models (\BothTen{} and \DMTen{}) show stronger effects of the scalar field in  dark matter power spectra than in  gas power spectra. This is because the power spectra of a component are most sensitive to changes to that  component. The smaller differences between the various models in the gas spectra are due to the simulation, which  is strongly dominated by dark matter compared to the gas, so that the extreme gas model (\GasTen{}) is  suppressed by  dark matter. This  also accounts for that fact that \GasTen{} is not  clearly distinguishable in  dark matter spectra.

On smaller scales, $k>1$ Mpc/h, the variations between the models are most noticeable in  gas spectra, as is to be expected as the baryonic processes are strong at the smaller scales.
Overall, we see a clear example that big deviations in  dark matter power spectra do not mean big deviations in the gas power spectra.

\subsection{Power spectra deviations}
The main difference between \BothTen{} and \DMTen{} in  dark matter power spectra deviations is that the latter is shifted slightly towards smaller scales and has a slightly smaller amplitude. The addition of the extremely coupled gas increases clustering on larger scales, while decreasing clustering on smaller scales. 

The gas spectra deviations show that the spectra is greatly influenced by both the dark matter and  gas. This is evident because \BothTen{} has the strongest deviations, while the second and third largest deviations come from \GasTen{} and \DMTen,{} respectively. The effect of dark matter on the gas spectra is much stronger than the effect of the gas on the dark matter spectra, a consequence of  the dominance of dark matter.

\BothOne{} and \GasPointOne{} show that  a minimally coupled gas reduces the amplitude of the power spectra in the range $k\in[0.1,3)$, with the largest difference from \BothOne{} at $k\sim1$ Mpc/h. However, at the smallest scales these models end up with the same amplitude, indicating that at this point the gas coupling is irrelevant for  dark matter spectra.

The models \BothPointOne{} and \DMPointOne{} reveal that the gas has an impact on  dark matter spectra, although not significantly, and the same very minor effect was also evident when comparing \GasTen{} to \BothTen{} and \DMTen{}.

The gas spectra is more susceptible to changes to  dark matter coupling, however, it is still most dependent on  gas coupling, as demonstrated by the increasing amplitude as \DMTen{}$<$\GasTen{}$<$\BothTen{}. All the models with an extreme coupling have  very low amplitude at small scales, and at the very smallest scales, they all end up at more or less the same value regardless of whether the coupling is universal or not. The same effect is seen when comparing \BothOne{} to \DMPointOne{}.


\begin{figure}
        \centering
        \includegraphics[width=0.48\textwidth]{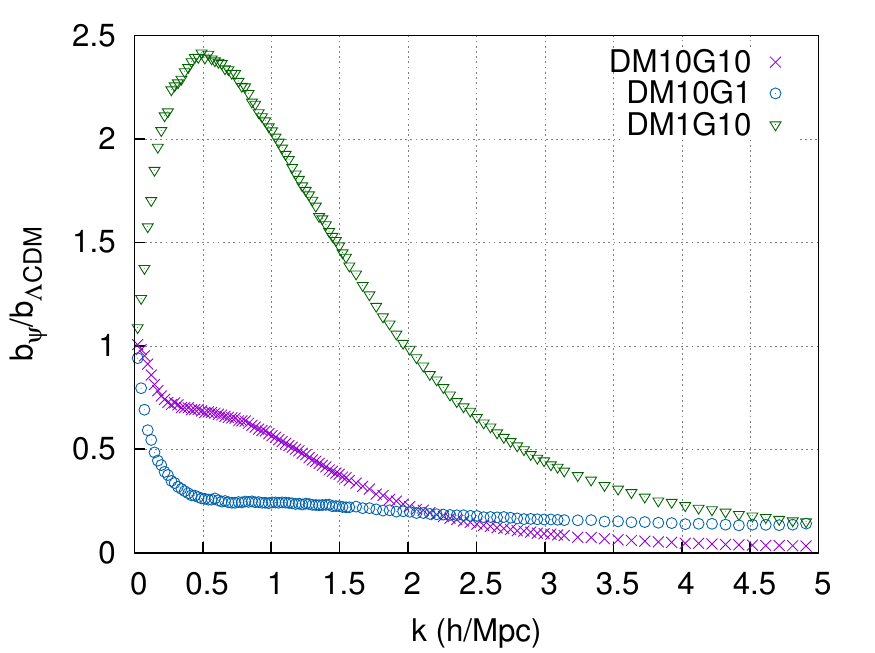}

        \includegraphics[width=0.48\textwidth]{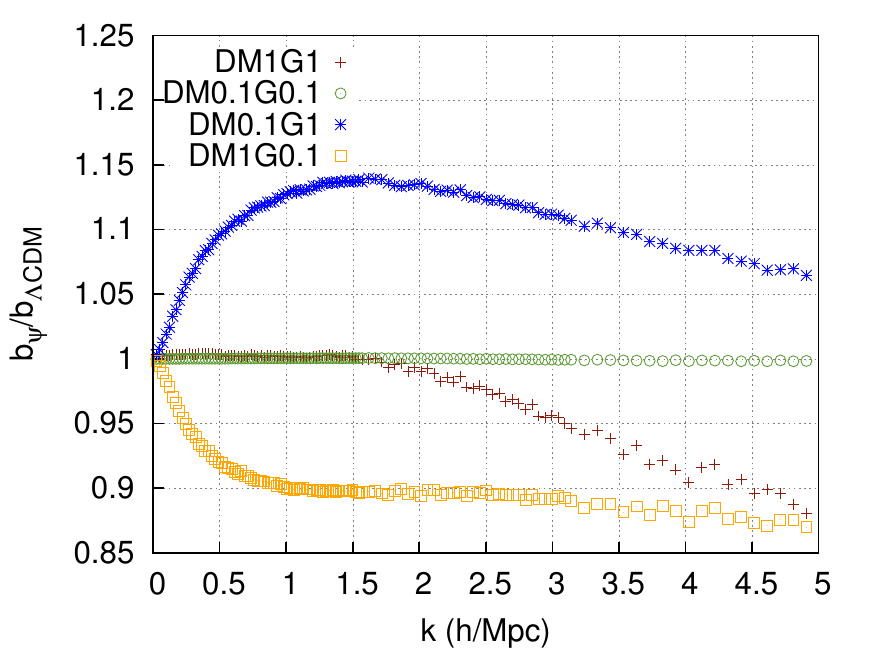}
        \vspace{-1 mm}
\caption{\label{biasSym} Deviation of the bias from the $\Lambda$CDM bias for our various models. The top image shows the extreme models, the bottom image the remaining models. All results are at $z=0$.}
\end{figure}

\subsection{Bias}
The bias is defined as the ratio between the gas power spectrum and the dark matter power spectrum, 
\begin{align}
 b = \frac{P_{\rm Gas}}{P_{\rm DM}},
\end{align}
and is shown in \fig{biasSym}.
In the very large scale region, $k<0.3$ Mpc/h, the \GasTen{} model has a bias that sky rockets to values larger than the $\Lambda$CDM bias. On the other hand, \DMTen{}  exhibits the exact opposite behaviour with a lower bias that rapidly decreases. \BothTen{} displays behaviour intermediate between  these two models, but much closer to \DMTen{}. 

The remaining large scales, $0.3<k<1$ Mpc/h, prove a turning point for all models. \GasTen{} stops its growing bias and starts to decrease, and \DMTen{} slows its rapid decrease and seems to stabilise at a constant value with $b_{\psi}=0.25b_{\Lambda \rm CDM}$. Similarly, \BothTen{} starts to level out, however, as \GasTen{} starts to decrease \BothTen{} does as well.

On smaller scales, the biases of these models start to diminish and eventually all are at lower values than the $\Lambda$CDM bias. Bias deviations in the smaller, non-linear scales are strongly correlated with  gas coupling, while also dependent on the coupling to the dark matter. \BothTen{} reveals that the components compound the effect of the deviations so that the \BothTen{} deviation is larger than the sum of the \DMTen{} and \GasTen{} deviations.
\BothPointOne{} exhibits almost no deviations from $\Lambda$CDM, as is to be expected. In the final models, the behaviour of the bias deviations act in extremely different manners depending on what components are strongly coupled.

For \GasPointOne,{} we see that the bias deviations plummet at  large, non-linear scales until they reach more or less constant values that are decreasing slightly from $k\sim1$ Mpc/h and out. The \DMPointOne{} model exhibits the exact opposite behaviour at  large, non-linear scales. \BothOne{} displays a behaviour intermediate between the other two models at the large scales, and then eventually starts to decrease to much less than the sum of the respective deviations.
The power spectrum bias at larger scales shows that models with a universal coupling have fewer deviations from the $\Lambda$CDM model than the models with a non-universal coupling.

For observational astronomers the implications of this behaviour is that if the bias deviations are greater than unity, $b_{\psi}>b_{\Lambda \rm CDM}$, then researchers who infer dark matter  properties from baryonic physics features will make predictions with values of the power spectrum that are too high. The opposite is true if the bias deviations are less than unity, $b_{\psi}<b_{\Lambda \rm CDM}$.

\begin{figure*}
%
%
\hspace{2.75cm}Dark matter \hspace{5.3cm}\;Gas\\
\vspace{-2mm}
        \centering
        \includegraphics[width=0.4\textwidth]{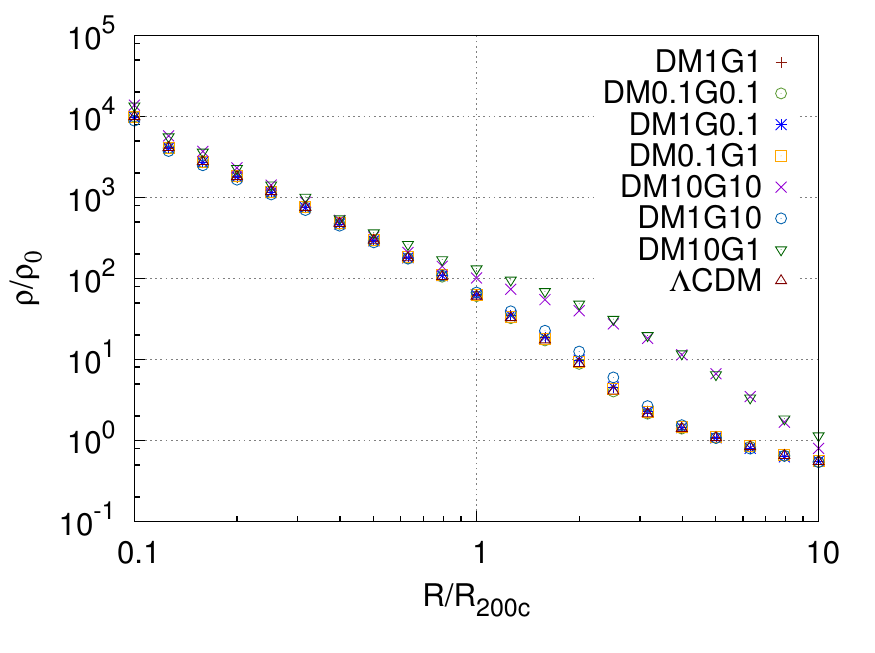}
        \vspace{-4 mm}
        \hspace{-5 mm}
        \includegraphics[width=0.4\textwidth]{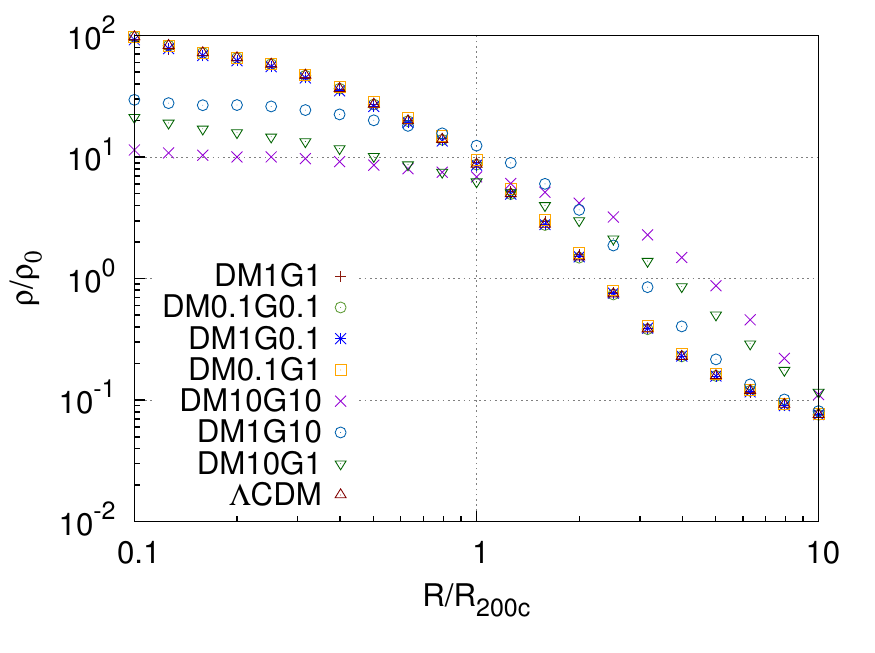}

        \includegraphics[width=0.4\textwidth]{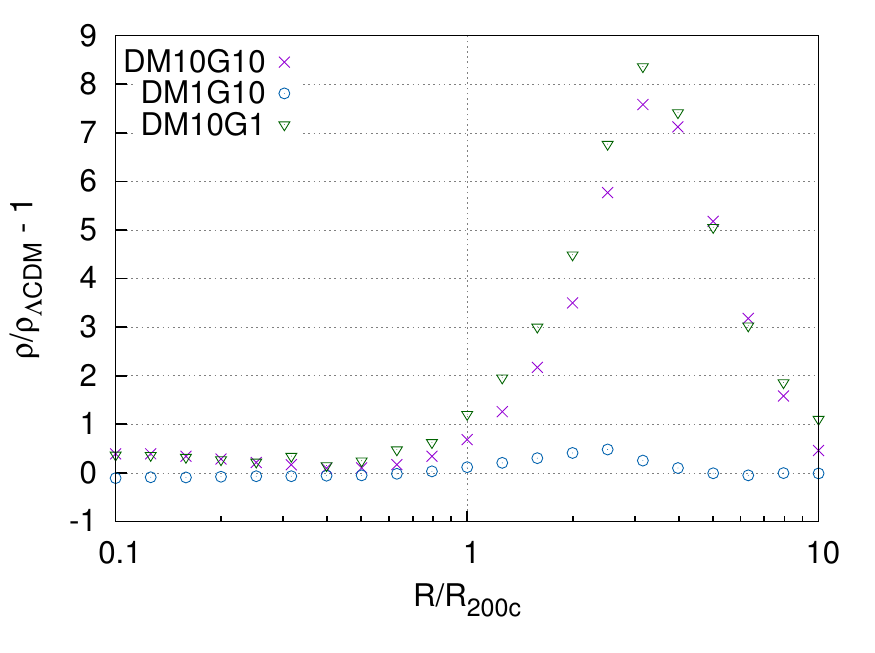}
        \vspace{-5 mm}
        \hspace{-5 mm}
        \includegraphics[width=0.4\textwidth]{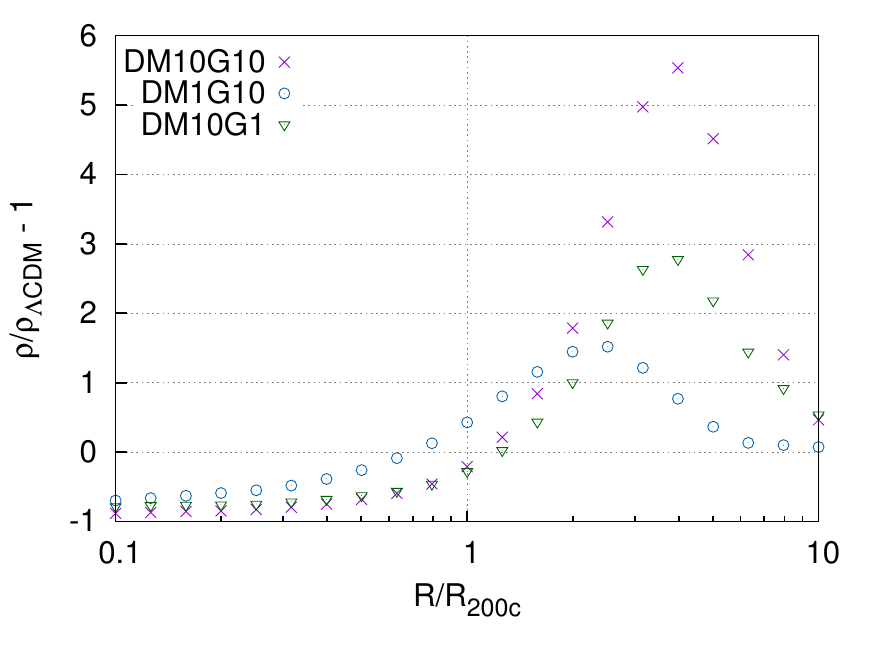}

        \includegraphics[width=0.4\textwidth]{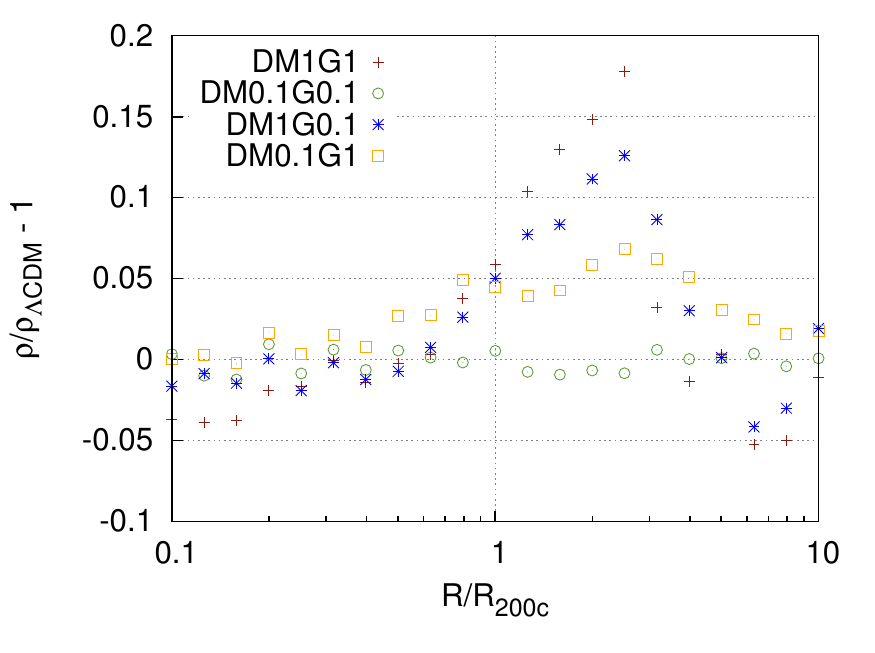}
        \vspace{-5 mm}
        \hspace{-5 mm}
        \includegraphics[width=0.4\textwidth]{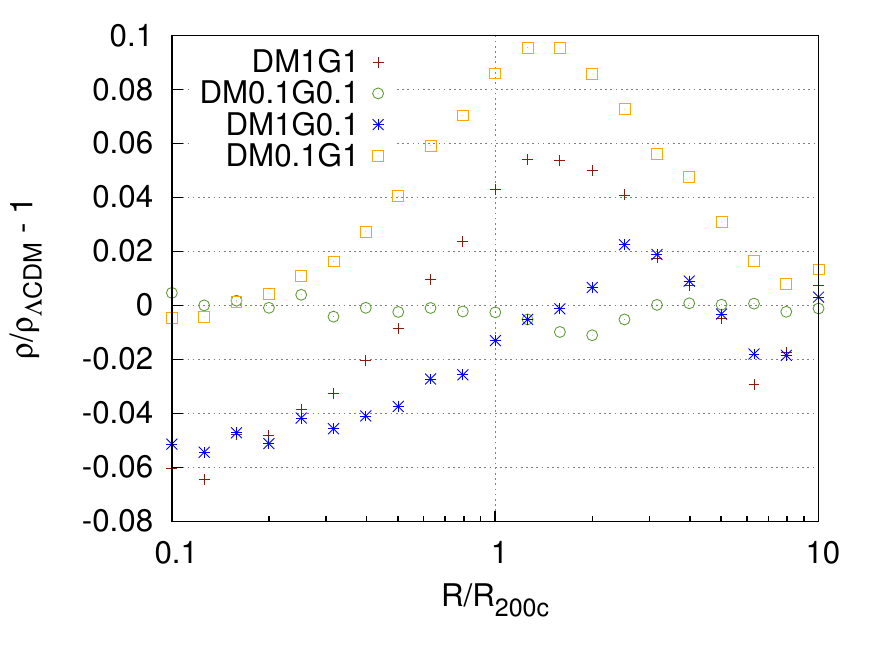}

\caption{Density properties for the massive halos with mass in the range [1$\times10^{14}h^{-1}M_{\odot}$,$5\times10^{14}h^{-1}M_{\odot}$). Top: density profiles for $\Lambda$CDM and  symmetron models. Middle: density profile deviations from $\Lambda$CDM for the extreme models. Bottom: density profile deviations from $\Lambda$CDM for the remaining models. Left column shows the dark matter component, while the right column shows the gas component. All results are at $z=0$.}
 \label{fig:Symmetron_density_profiles}
 \end{figure*}

\section{Halo profiles}
\label{profile_results}
In this section, we present density and temperature profiles for multiple halos identified by using the  Rockstar code developed by \citet{rockstar}. We study massive halos with mass in the range [1$\times10^{14}h^{-1}M_{\odot}$,$5\times10^{14}h^{-1}M_{\odot}$), the high mass of these halos should ensure that the screening mechanism are triggered in dense regions.

In previous works \citep{hydro_mod, ISIS}, the halos that had not yet reached a relaxed state were filtered out  following the methods described in \citet{relax} and \citet{Shaw}. The method used relations between the kinetic and potential energy and  surface pressure to determine if a halo was relaxed or not. \citet{symmfofrredshift} refined the method to take the effects of modified gravity in the virialisation state of the halos into account. However, if we filtered the non-relaxed halos, this would eliminate too many halos in the extreme models and leave almost no remaining halos. For this reason, we do not filter the non-relaxed halos from the relaxed halos. 

The halo profiles are calculated by sweeping over all the cells in the simulations, binning dark matter particles and  baryonic gas properties in annular bins for each halo\footnote{Rockstar filters out unbound particles when calculating the halo properties, while we construct the profiles by sweeping over all cells and binning particles within $r=10R_{\rm 200c}$, potentially including these unbound elements.}, then averaging over all halos. The profiles range from 10\% of the virialisation radius, $r=0.1R_{\rm 200c}$, to ten times the virialisation radius, $r=10R_{\rm 200c}$.  This range was chosen to properly catch all behaviours of the fifth force on the dark matter and gas halos, while also avoiding the inner regions of the halos where the resolution of our simulations is low.  All profiles are calculated at the present epoch $z=0$.

\subsection{Density profiles}

The density halo properties are presented in \fig{fig:Symmetron_density_profiles}. An extreme coupling in the dark matter  strongly affects both   dark matter and gas density profiles. For  dark matter profiles, the effect is mostly contained at the exterior of  dark matter halos, while  the gas profiles show signs of \ extreme dark matter coupling in all parts of the halo and outside. The diminished clustering at the inner regions of the halos can be explained with an environmental effect from the dark matter on the gas.

The dark matter   clusters faster than the gas because of its collisionless nature, in which the gas is prevented from collapsing, due to the pressure, and the dark matter clusters unhindered. This means that the dark matter  reaches higher densities at a faster rate than the gas, and from the description of the screening mechanism, we know that the screening is triggered by a combined density, dark matter plus gas, threshold, resulting in the dark matter triggering the screening mechanism before the scalar field has had a chance to work on the gas component as much as on the dark matter. 

An extremely coupled gas has a minor, however, not negligible effect on the dark matter profiles, while it has a huge impact on the gas profiles. This is expected as there is much more dark matter than gas, resulting in the effect of the gas not being as strong. Also,  \BothTen{} is the model with the biggest effect on the gas profiles, while the dark matter profiles seem to prefer \DMTen{}.

\subsection{Density profiles deviations}
The deviations from $\Lambda$CDM for the halos are found in the lower two rows of \fig{fig:Symmetron_density_profiles}. The deviations confirm the conclusions from the total density profiles, and also reveal previously unseen effects. All models without extremely coupled dark matter make dark matter halos cluster less  at the inner regions than in those two cases   (\BothTen{} and \DMTen). The exact opposite occurs in the gas power spectra, where all extremely coupled models (the two previously mentioned and \GasTen) cluster less than the other models. This is due to  baryonic physics preventing the gas from collapsing as far inwards as the dark matter, and then the gravitational contribution from the gas on the dark matter also prevents the dark matter from clustering.

The peak of deviations does not occur at the same radius for all the models. For \GasTen,{} the peak in deviations is closer to the halo centre, while for \BothTen{} and \DMTen{} the deviations peak further out. This effect is due to the extreme coupling in the gas. The extreme coupling allows the scalar field to affect the gas component to such an extent that it collapses further inwards before the environmental effect of the combined density triggers the screening mechanism.

The bottom row reveals that \GasTen{} has a strong effect on  dark matter halos throughout the  density profiles with higher peak deviation than \BothOne{}. The gas profiles are under-dense at the inner regions of the halos in the cases where the dark matter is normally (or extremely) coupled to the scalar field. In  the cases where the dark matter is minimally coupled, the profiles approach $\Lambda$CDM at the centre of the halos. This behaviour further asserts our conclusions about the environmental effect.


To study how deviations from $\Lambda$CDM differ in  dark matter and gas cases, we introduce the deviation bias $\delta^{\rm DM}$, defined as the relative difference between the deviations
\begin{align}
 \delta^{\rm DM} \!=\! \frac{\Delta_{\rm DM}-\Delta_{\rm Gas}}{\Delta_{\rm Gas}} \!=\! \dfrac{\frac{\rho_{\rm DM}-\rho_{\Lambda \rm CDM}}{\rho_{\Lambda \rm CDM}} - \frac{\rho_{\rm Gas}-\rho_{\Lambda \rm Gas}}{\rho_{\Lambda \rm Gas}}}{\frac{\rho_{\rm Gas}-\rho_{\Lambda \rm Gas}}{\rho_{\Lambda \rm Gas}}}\!,\!\!\!
\end{align}
where $\rho_{\Lambda \rm Gas}$ is the gas $\Lambda$CDM density and $\rho_{\Lambda \rm CDM}$ is the DM $\Lambda$CDM density.
 
\begin{figure}
        \centering
        \includegraphics[width=0.48\textwidth]{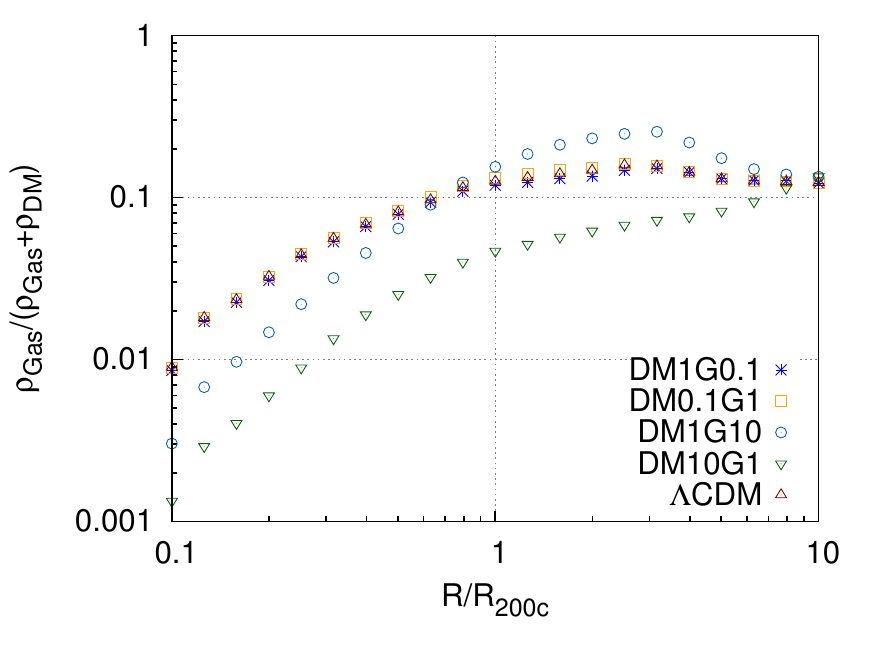}
        \vspace{-5 mm}
        \hspace{-5 mm}
        \includegraphics[width=0.48\textwidth]{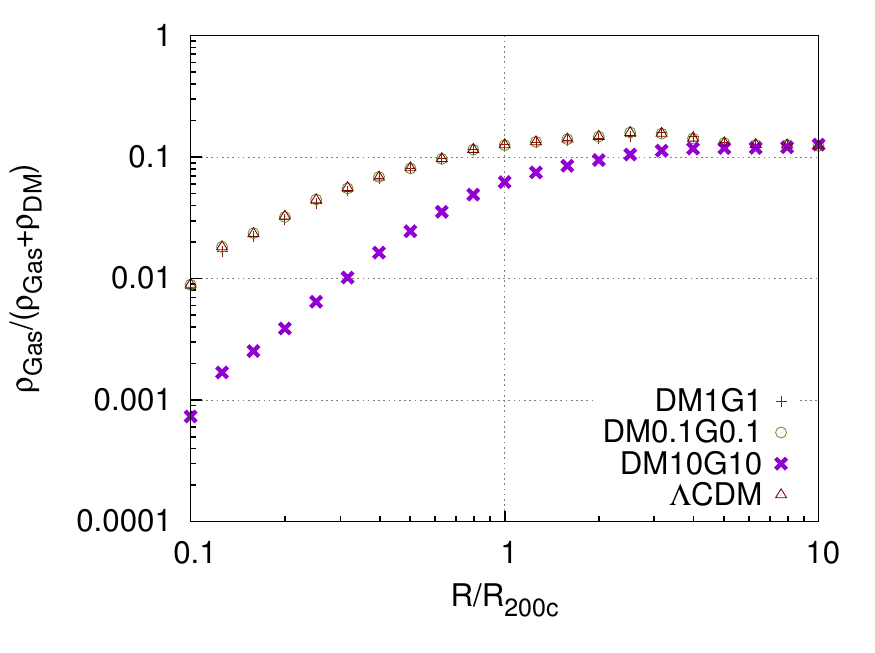}
        \vspace{-5 mm}
\caption{\label{dens_bias} Baryon fraction profiles for the halos with mass in the range [1$\times10^{14}h^{-1}M_{\odot}$,$5\times10^{14}h^{-1}M_{\odot}$). Top: models with non-universal coupling. Bottom: models with universal coupling. All results are at $z=0$.}
\end{figure}

The \DMTen{} model represents a model that has an enormous deviation from $\Lambda$CDM in the dark matter component of $\Delta_{\rm DM}\sim8.5$ at $R \sim4R_{\rm 200c}$, and a much smaller deviation of $\Delta_{\rm Gas}\sim2.5$ in the gas component, giving a deviation bias of $\delta^{\rm DM}\approx 2.4$. The same effect, but now with enourmous deviations in the gas component, is present in \GasTen{} where the deviations from $\Lambda$CDM are $\Delta_{\rm Gas}\sim1.5$ for the gas and $\Delta_{\rm DM}\sim0.5$ for the dark matter, giving a deviation ratio of $\delta^{\rm DM}\approx -\frac{2}{3}$.

\subsubsection{Baryon fraction profiles}
One of our aims is to find a method of distinguishing between models that have universal coupling and models with non-universal coupling. For this purpose, we use the baryon fraction \citep{gasfrac},
\begin{align}
f_{\rm Gas} = \frac{\rho_{\rm Gas}}{\rho_{\rm Gas}+\rho_{\rm DM}},  
\end{align}
and present the baryon fraction profiles in \fig{dens_bias}.

The baryon fraction shows that all the models have the same behaviour as $\Lambda$CDM, except the three models \BothTen{}, \GasTen,{} and \DMTen{}. \BothTen{} converges with the $\Lambda$CDM baryon fraction earlier than \GasTen{} and \DMTen{}, however, in general this model has a higher deviation at the inner region.  In fact, the models with a universal coupling seem to deviate less from $\Lambda$CDM than the models with a non-universal coupling.

This is the same behaviour displayed by the power spectrum bias at larger scales, and allows us to conclude that significant deviations in the baryon fraction from the $\Lambda$CDM model throughout the halos might be an indication of a non-universal coupling.

\begin{figure}
        \centering
        \includegraphics[width=0.4\textwidth]{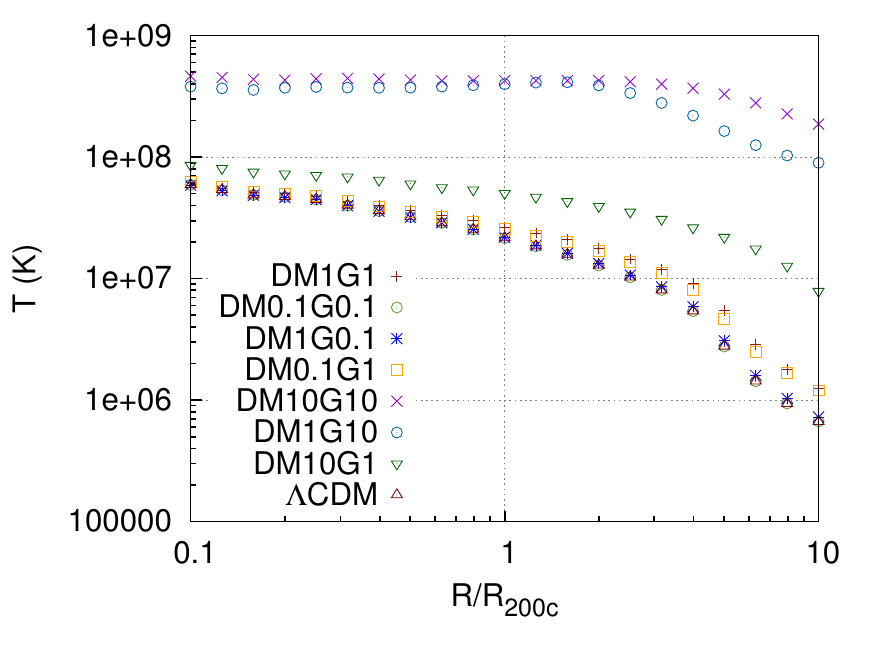}

        \includegraphics[width=0.4\textwidth]{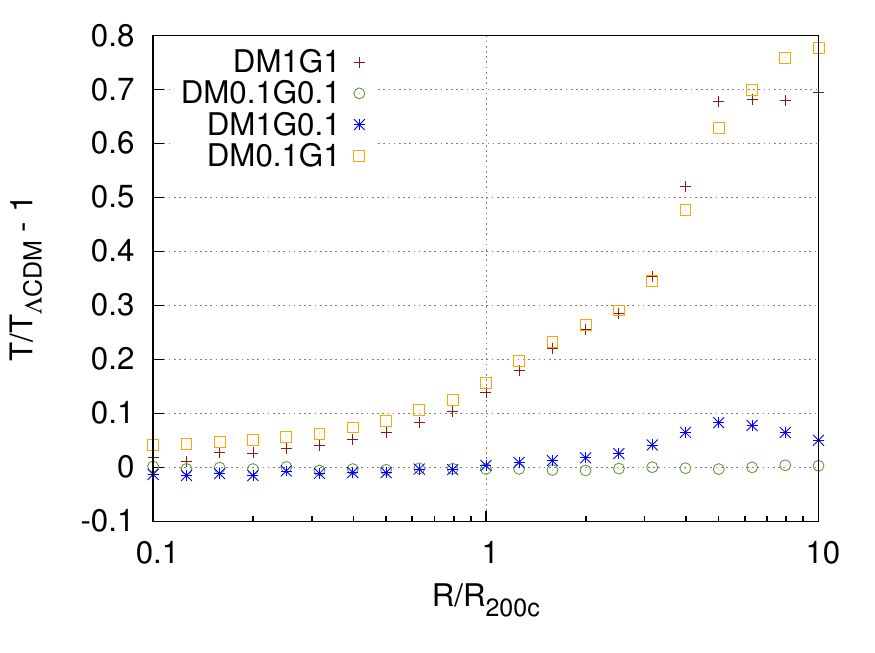}

        \includegraphics[width=0.4\textwidth]{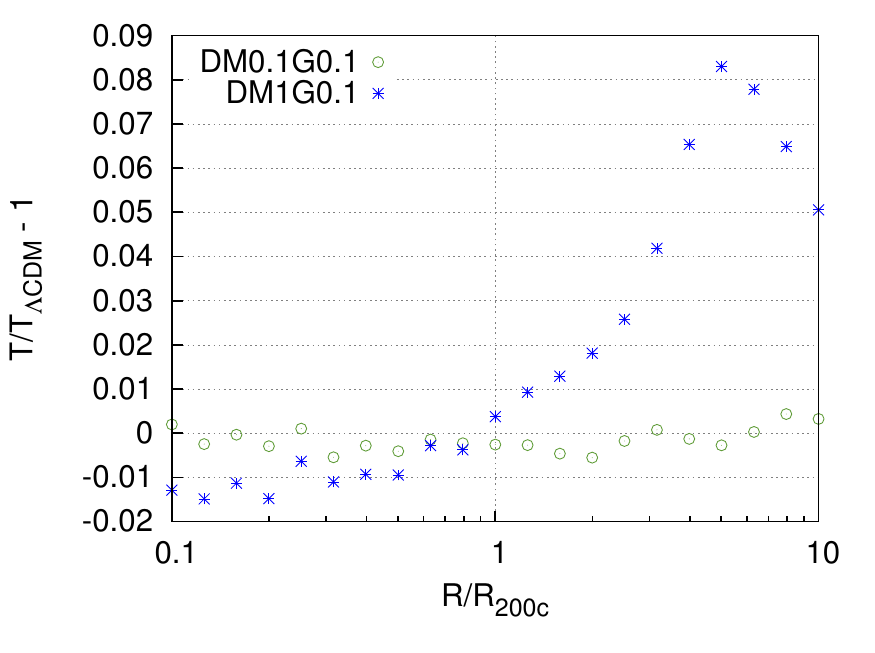}

        \vspace{-1 mm}

\caption{Temperature properties for the our halos. Top: temperature profiles for $\Lambda$CDM and  symmetron models. Middle: temperature profile deviations from $\Lambda$CDM. Bottom: deviations with the extremely coupled models filtered out. All results are at $z=0$.}
 \label{fig:temperature_profiles}
 \end{figure}

\subsection{Temperature}
\label{temp_prof_results}
The temperature is a very interesting component to study due to its close relation observables \citep{XMM, TestCham}. The temperature is not an output of our code and needs to be reconstructed using the ideal gas law,
\begin{align}
p = R_{\rm s}\rho T,
\end{align}
where $p$ is the thermal pressure, $R_{\rm s}=\frac{k_B}{\bar{m}}$ is the specific gas constant, $\bar{m}=0.59m_H$ is the mean mass of the gas, $m_H$ is the hydrogen mass, and $\rho$ is the gas density. The temperature profiles are made and analysed in the exact same manner as the density profiles.

\subsubsection{Total temperature profiles}
The temperature profiles are presented in \fig{fig:temperature_profiles}. The temperature is a product of the baryonic processes and is most sensitive to strong coupling between the gas and scalar field \citep{hydro_mod}. 

The stronger the coupling to the scalar field is, the higher the temperature in the halos is. The slope of the profiles outside of the halo, as the temperature starts dropping, is shallower for \BothTen{}, \GasTen,{} and \DMTen{} than in the other models.  \BothTen{} gives an even higher temperature than \GasTen, indicating that the dark matter plays a vital role in the temperature of baryonic halos. This is also evident from \DMTen, which has a higher temperature than the less extreme cases.

\subsubsection{Temperature profiles deviations}
The temperature deviations from the $\Lambda$CDM temperature are found in the bottom two rows of \fig{fig:temperature_profiles}.
\DMPointOne{} and \BothPointOne{} are two similar models in which one has universal coupling and the other does not. These two models display no signatures that can be interpreted as a trait of non-universal coupling. \GasPointOne{} and \BothPointOne,{} on the other hand, display an increase in temperature outside of the virialisation radius.

Both of these effects  come from the fact that these halos are so massive that they are most likely screened from the scalar field. The difference between \GasPointOne{} and \BothPointOne{} is that the regions outside of the halos are non-screened, and the stronger coupling to dark matter allows \GasPointOne{} to be influenced by the scalar field to a larger extent than \BothPointOne{}.

\section{Conclusions}
\label{discuss}
We investigate scalar-tensor theories of gravity, which present a non-universal coupling. That is, baryons and dark matter are coupled with the scalar degree of freedom with different coupling strengths. The models we investigate utilise a screening mechanism to suppress the deviations from GR at small (solar system) and large cosmological scales:  the symmetron screening mechanism, specifically.

As a result of the screening mechanism, the strongest signatures in these models are expected to occur at the non-linear regime of structure formation. Therefore, in order to unveil the imprints of these theories at astrophysical scales, we ran several hydrodynamic cosmological N-body simulations. We compared models with and without a universal coupling to the symmetron scalar field, and showed that several astrophysical observables (density profiles, temperature profiles and power spectra) show significant differences between the dark matter and gas components when the coupling is non-universal.

The deviations from $\Lambda$CDM are typically larger in the gas than in the dark matter near the centre of the halo. The opposite holds true at larger radii, where the dark matter deviates more strongly from $\Lambda$CDM. However, this is not the case for  models in which coupling of the gas is significantly stronger than that of the dark matter.


For  power spectra,  dark matter deviations are larger than that of the gas in models with universal coupling or in  models in which  dark matter coupling is stronger than  gas coupling. 


Our attempt to find signatures in  density profiles and power spectra, which would reveal  whether  coupling to the scalar field is universal or not, revealed one signature: in the cases of universal coupling, the deviations in the baryon fraction and bias from $\Lambda$CDM are smaller than in the cases of non-universal couplings throughout the halos. This is expected, since GR is a universally coupled theory. If observers find the baryon fraction or power spectrum bias to deviate from the calculated $\Lambda$CDM bias, then this might very well be a sign of a non-universal coupling, and therefore a breaking of the equivalence principle.

Separating the dark matter and gas will prove to be a challenge for observers intending to compare their results with theories, since the dark matter is not a direct observable, while the gas is observable.   To work around this, we propose to use the baryon fraction as presented in this paper. To measure this baryon fraction, first we suggest the constructing of the total density profile, $\rho_{\rm Tot} = \rho_{\rm DM}+\rho_{\rm gas}$,  using rotational velocity profiles of individual galaxies to measure the dynamical mass, and then constructing the gas density profile $\rho_{\rm gas}$,  using X-ray surface brightness and X-ray temperature profiles.

A caveat with the above mentioned method is that the calculated dynamical mass is dependent on the particular gravity model used. An alternative option is to calculate the lensing mass using gravitational lenses as the path of light that is independent for all conformal scalar gravity theories \citep{Bekenstein,SeanMCarroll} and generally ideal for studying modified gravity theories \citep{lensmass}.

The above method for detecting non-universal couplings may not be possible with the current state of observational and theoretical limits. The deviations we found  from $\Lambda$CDM are all quite small, when we exclude the extreme models, and also contain large uncertainties depending on the modelling of feedback physics. 

Furthermore the observational baryon census of galaxies contains significant errors and may still be incomplete, i.e. the halo missing baryon problem \citep{Werk}, while the baryon consensus in galaxy clusters is less challenging as the majority of the gas is hot enough to be visible in X-rays. The above method is therefore more likely, yet still very challenging, to work in galaxy structures than in galaxies.

If one takes seriously the possibility of matter components with a non-universal coupling to a gravity scalar degree of freedom, then our work shows the bias will be greatly affected. Therefore, attempts to rule out or constrain modified gravity theories by comparing dark matter predictions to the observed quantities based on baryonic properties may be misleading, and one must consider the possibility of a non-universal coupling that might skew the conclusions and  dark matter properties that are inferred from baryons.

\section{Acknowledgements}
The authors thank C. Llineares and H. Winther for sharing the ISIS code to run the simulations performed in this article. We also thank M. Grönke and H. Winther for comments and suggestions. The authors thank The Research Council of Norway for funding and the NOTUR facilities for the Computational resources. 

\bibliography{Diffcop_bib}

\end{document}